\documentclass[12pt,]{article}

\usepackage{amsmath,amssymb,epsf}
\usepackage{graphicx,color}
\usepackage{epstopdf}
\usepackage[normalem]{ulem}
\usepackage{color}
\def\bel#1{\begin{equation} \label{#1}}

\def\be{\begin{equation}}
\def\ee{\end{equation}}
\def\bea{\begin{eqnarray}}
\def\eea{\end{eqnarray}}

\def\ltap{\ \raise.3ex\hbox{$<$\kern-.75em\lower1ex\hbox{$\sim$}}\ }
\def\gtap{\ \raise.3ex\hbox{$>$\kern-.75em\lower1ex\hbox{$\sim$}}\ }
\def\gl{\ \raise.5ex\hbox{$>$}\kern-.8em\lower.5ex\hbox{$<$}\ }
\def\roughly#1{\raise.3ex\hbox{$#1$\kern-.75em\lower1ex\hbox{$\sim$}}}

\newcommand{\comments}[1]{}

\newcounter{oldcounter}
\addtocounter{equation}{1}
\begin{document}

\begin{titlepage}
\vskip 1 cm
\begin{center}
{\Large \bf   Curvature Singularity in $f(R)$ Theories of Gravity
}
\vskip 1.5cm  

{ 
\sc{Koushik Dutta$^{*}$, Sukanta Panda$^{\dagger}$, Avani Patel$^{\dagger}$}
\let\thefootnote\relax\footnotetext{ \hspace{-0.5cm} E-mail: {$\mathtt{koushik.dutta@saha.ac.in, sukanta@iiserb.ac.in, avani@iiserb.ac.in} $}}
}
\vskip 0.9 cm

{\textsl{$^{*}$Theory Division, \\
 Saha Institute of Nuclear Physics, \\
 1/AF Salt Lake,  \\
 Kolkata - 700064, India}
}
\vskip 0.6 cm
{\textsl{
$^{\dagger}$IISER, Bhopal,\\ Bhauri, Bhopal 462066,\\
Madhya Pradesh, India\\}
}
%
%
\end{center}

\vskip 0.6cm

\begin{abstract}
\vskip 0.5 cm
{
 Although $f(R)$ modification of late time cosmology is successful in explaining present cosmic acceleration, it is difficult to  satisfy the fifth-force constraint simultaneously. Even when the fifth-force constraint is satisfied, the effective scalar degree of freedom may move to a point (close to its potential minima) in the field space where the Ricci scalar diverges. We elucidate this point further with a specific example of $f(R)$ gravity that incorporates several viable $f(R)$ gravity models in the literature. In particular, we show that the nonlinear evolution of the scalar field in pressureless contracting dust can easily lead to the curvature singularity, making this theory unviable.}
\noindent

\end{abstract}

\vspace{3.0cm}

\end{titlepage}



\section{Introduction}
Modern cosmological models suffer from a major theoretical difficulty, namely, the dark energy problem. The problem stems from the high-precision observational data with strong evidence that the universe is undergoing a phase of accelerated expansion in recent times \cite{Nobel2011}. Many models have been constructed for an explanation of the late time acceleration of the Universe, and those models can be mainly divided into two categories: dark energy models that change the matter content of the Universe and modified gravity models that alter the Einstein gravity. 

Dark energy is generally modelled by the vacuum energy with an equation of state parameter $ w= -1$ or by a dynamical scalar field (dubbed as quintessence) with $w \simeq -1$ \cite{Fujii82}. Even though both the cosmological constant and the quintessence field are observationally consistent in explaining the cosmic acceleration, the origin of these matter sources from fundamental physics is not well-understood \cite{Padilla:2015aaa}. 
  
On the other hand, late time acceleration can also be obtained from the modifications of General Relativity, which introduces an extra degree of freedom to the gravitational sector itself. The simplest model in this category are $f(R)$ theories, where $f(R)$ is an arbitrary (usually analytic) function of $R$ with $R$ being the Ricci scalar \cite{f(R)TheoriesReview, Capozziello:2011et, Sotiriou:2008rp, Nojiri:2010wj}. The equation of motion of the extra degree of freedom other than the graviton is of second order \cite{Woodard:2006nt}, and the models are free from classical and quantum instabilities \cite{Dolgov:2003px, Starobinsky:1980te}. One stringent constraint on these models is imposed from the avoidance of fifth force carried by the extra scalar degree of freedom \cite{Will:2014xja, Khoury:2003aq, Capozziello:2007eu}, and a few successful models have been constructed \cite{Starobinsky:2007hu, Hu:2007nk, Appleby:2007vb}. These models produce observationally consistent accelerating expansion preceded by the matter domination \cite{Amendola:2006we}. 

Lately, the curvature singularity occurring at cosmological time scales have been found to be a serious problem in $f(R)$ models \cite{Frolov:2008uf}. It is well known that $f(R)$ theories can be considered as equivalent scalar-tensor theories. It is easy to visualise curvature singularity in a $f(R)$ theory by looking at the form of the potential appearing in its equivalent scalar-tensor theory in the Jordan frame \cite{Dev:2008rx}. Due to the nonlinear motion of the scalar field, the oscillations around the potential minimum can make the field displaced to the singular point. The presence of the matter makes the occurrence of singularity more probable \cite{Frolov:2008uf}. The finite-time singularity in modified gravity is described in \cite{Nojiri:2008fk, Bamba:2008ut}, whereas the singular behaviour of curvature in a contracting universe has been analysed in  \cite{Appleby:2008tv}. It is shown that past singularities may be prevented for a certain range of parameters. These singularities may also occur in future and can be avoided for fine-tuned initial conditions \cite{Dev:2008rx, Thongkool:2009js}. It is also realised that the curvature singularities can be eliminated by adding an extra curvature term to the Lagrangian \cite{Appleby:2008tv, Appleby:2009uf}. The curvature singularity can also be seen in an astrophysical object. In this case, the singularity is analysed for suitable $f(R)$ models applied to dense objects undergoing contraction in the presence of linearly time-dependent mass density \cite{Arbuzova:2010iu, Lee:2012dk}. It is seen that the singularity is reached in a time that is much shorter than cosmological time scale. A detailed study of this issue can be found in \cite{Reverberi:2012ew}.

In this work, we study the issue of curvature singularity in a $f(R)$ model proposed in \cite{Miranda:2009rs}. We give special emphasis on the role of external matter that makes the effective potential shallower compared to the pure vacuum case. We find that it is impossible to find a parameter space where both the fifth-force constraint and the curvature singularity issues are resolved. This paper is organised as follows: In Sec.~\ref{sec:gendiscussion}, we review the curvature singularity and fifth force constraints for a general function $f(R)$. The occurrence of curvature singularities and fifth force constraint in a specific model of \cite{Miranda:2009rs} is analysed in Sec~\ref{sec:genmodel}. We extend the same analysis to a specific limit of the above mentioned  model in Sec.~\ref{sec:logmodel}. Finally, we summarise our results in Sec.~\ref{sec:conclusion}.

\section{Curvature Singularity and Fifth-Force Constraints in $f(R)$ Gravity}\label{sec:gendiscussion}
In $f(R)$ theories of gravity, the Einstein-Hilbert action is modified by replacing the Ricci scalar with an arbitrary function of $R$,
\begin{equation}
S=\frac{1}{2 \kappa^2}\int d^4x\sqrt{-g} f(R)+ S_M
\label{m1action}
\end{equation}
with $\kappa^2 = 8\pi G = M_{Pl}^{-2}$, where $S_M$ contains the matter degrees of freedom which does not include contributions from dark energy. 
For our convenience, we will write the arbitrary function in the following form $f(R) = R+ F(R)$, so that $F(R)$ captures the modifications of Einstein gravity. The effect of $F(R)$ in the cosmological dynamics is relevant when local curvature becomes smaller than a characteristic infrared modification scale $R_{*}$. Above this scale, the gravity behaves approximately as Einstein gravity. 

By varying the action with respect to the metric tensor $g_{\mu\nu}$ one can obtain the field equation, and taking trace of that field equation we arrive at 
\begin{equation}
3\Box F_{,R}(R) -2F - R + R F_{,R}(R) = \kappa^2 T~,
\label{trace} 
\end{equation}
where $T$ is the trace of the stress-energy tensor coming from the matter Lagrangian, and a comma with the subscript corresponds to the partial derivative with that quantity. 
The scalar-tensor representation of $f(R)$ theory can be a found by identifying the term $F_{,R}$ as a dynamical degree of freedom, and in literature, it is dubbed as the `scalaron' field $\phi=F_{,R} = f_{,R}-1$ \cite{Starobinsky:1980te}. The scalar curvature of the space-time is governed by the local value of the field $\phi$. 

The above trace Eq.~\eqref{trace} can be conveniently written as 
\begin{equation}
\Box \phi = \frac{dV_J}{d\phi} + \frac{\kappa^2}{3} T~,
\label{wave}
\end{equation}
where
\begin{equation}
\frac{dV_J}{d\phi}=\frac{1}{3}(R+2F-RF_{,R})~.
\label{eq1fr}
\end{equation}
It may not be always possible to invert the relation $\phi=F_{,R}$ and obtain $R(\phi)$. Therefore it is convenient to write $V_J$ in a parametric form
\begin{equation}
\frac{dV_J}{dR} = \frac{dV_J}{d\phi} \frac{d\phi}{d R} = \frac{1}{3}(R + 2 F - RF_{,R}) F_{,RR}~.
\end{equation}
Here, the subscript $'J'$ denotes the potential in the Jordan frame. The dynamics of the field $\phi$ is governed by its potential $V_J$ originating from the modified gravity, and a force term that is proportional to the stress-energy tensor $T$. Integrating this expression we can find $V_J(R(\phi(x,t)))$. Thus, we can effectively describe the dynamics of modified gravity by the dynamics of a scalar field whose value is  uniquely determined by $f(R)$. 

\subsection {Curvature Singularity:} The potential $V_J(\phi)$ will have a global minimum at $\phi_{min}$ where cosmological evolution happens. But, as we will see, there is also a point in the field space, denoted by $\phi_{sing}$, where the curvature scalar $R$ diverges to infinity resulting into curvature singularity. Typically, the points $\phi_{min}$ and $\phi_{sing}$ would be separated by a finite field value with a finite energy barrier. Therefore, the field can potentially reach to the singular point in the process of having small oscillations around its minimum, in particular when we consider objects with growing mass density \cite{Frolov:2008uf}. The minimum of the potential $\phi_{min}$ can be obtained by equating $dV_J(\phi)/d\phi=R+2F-RF_{,R}=0$ which corresponds to a constant curvature solution $\Box F_{,R}=0$ for the vacuum. Thus, $\phi_{min}$ is also a de Sitter point; see Eq.\eqref{trace}. 

The appearance of the singularity can be seen both in the cosmological background \cite{Frolov:2008uf}, as well as in the astrophysical dense object going under spherical collapse \cite{Arbuzova:2010iu}. 
The Eq.\eqref{trace} can also simply be written as 
\begin{equation}
\Box \phi = \frac{d V_{J}^{\text{eff}}}{d \phi}~,
\end{equation}
where $ \frac{d V_{J}^{\text{eff}}}{d \phi} = \frac{d V_{J}}{d \phi} + (\kappa^2/3) T$, $i.e.$ $V_{J}^{\text{eff}}$ incorporates the effects of matter.  
Now, we would like to study the dynamics of the scalar field in an astrophysical system whose mass density increases with time, $i.e$ a collapsing object with time dependent $V_{J}^{\text{eff}}$. With the assumption of weak gravity, the covariant derivative can be replaced by the usual flat space ones \cite{Reverberi:2012ew}. Again, in the approximation of isotropic matter distribution (spherical collapse), spatial derivatives can be neglected \cite{Arbuzova:2010iu}. Thus, if we consider objects whose mass density changes with time $T = T(t)$ in a homogenous way, the above equation simplifies to
\begin{equation}
\frac{\partial^2 \phi}{\partial t^2} +  \frac{\partial V_J^{\text{eff}}(\phi,t)}{\partial \phi}  = 0~.
\label{diffaleq}
\end{equation}
This is an oscillator equation with a time dependent nonlinear potential. 
Even if the initial oscillation amplitude is small, due to the nonlinear behaviour of the motion, the above mentioned singular point can be reached during the evolution of the field. As we will see, in many viable $f(R)$ models when a collapsing astrophysical object is considered, $\phi$ evolves to the singular point in a finite time which is much shorter than the cosmological time scale $t_U \sim 4\times 10^{17}$ sec. This makes the theory not viable. 

\subsection{Fifth Force Constraint:} It is clear from the above discussion that the $f(R)$ gravity theory contains a scalar degree of freedom in addition to the usual graviton. The effect of the scalar degree of freedom in the matter sector is evident when the theory is rewritten in the Einstein frame, where the gravity part is Einstein Hilbert type. Taking the conformal transformation on the metric
\begin{equation}
\tilde g_{\mu\nu} = f_{,R}\; g_{\mu\nu}, \;\;\;\; \kappa\psi = \sqrt{3/2} \ln{f_{,R}}~,
\end{equation}
the action in \eqref{m1action} transforms to
\begin{equation}
S = \int {d^4x \sqrt{-\tilde g}\left[\frac{\tilde R}{2\kappa^2} - \frac{1}{2}(\tilde\nabla\psi)^2 - V_E(\psi) + \mathfrak{L}_m(\tilde g_{\mu\nu}e^{-\frac{2}{\sqrt{6}}\kappa\psi})\right]}
\label{Einsteinaction}
\end{equation}
where, $V_E$ is the scalar field potential in the Einstein frame
\begin{equation}
V_E = \frac{Rf_{,R}-f}{2\kappa^2f_{,R}^2}~.
\end{equation}
All the quantities having tilde are defined in Einstein frame. The mass of the scalar field $\psi$ can be as light as the present Hubble constant, and as a result, there will appear a long range fifth force mediated by the  field. In order to satisfy the local gravity constraints one must have some screening mechanism which can screen the fifth force on the surface of the Sun/Earth. One such mechanism is known as chameleon mechanism \cite{Khoury:2003aq}. 
The scalar field will be heavier in a dense object like earth, and light when the background density is less like the interstellar space. As long as the scalar field is heavy inside the surface of the earth, it will be frozen at the minimum of its effective potential
\begin{equation}
V_{\text{eff}}(\psi) = V_E(\psi) + e^{-\frac{2}{\sqrt6}\kappa\psi}\rho~,
\end{equation}
where $\rho$ is the energy density of the matter field and cannot contribute to the outside field. The only contribution to the outside field can be from a very thin shell around the surface of the earth. The thin shell parameter is given by \cite{Khoury:2003aq}
\begin{equation}
\frac{\Delta\tilde r_c}{\tilde r_c} = - \frac{\psi_{out}-\psi_{in}}{\sqrt{6}\Phi_{test}},
\label{thinshellcond}
\end{equation}
where $\psi_{in}$, $\psi_{out}$ are field values corresponding to the minimum of the effective potential inside and outside of a test body (Sun/Earth). $\Phi_{test}$ is the gravitational potential on the surface at the radius $\tilde r_c$ of the test body  and it is given by $\Phi_{test}=M_{test}/8\pi\tilde r_c$ where, $M_{test}=(4\pi/3)\tilde r_c \rho_{in}$. Here $\rho_{in}$ is the average density of the test body. $\Delta\tilde r_c$ is the width of the thin shell around the surface of the test body. In the usual approximation of $F_{,R}\ll1$ and $F\ll R$, it can be shown that $\psi_{in} \ll\psi_{out}$ for all practical purposes, and the thin-shell condition reduces to \cite{Will:2014xja}
\begin{equation}
|\psi_{out}| \lesssim \sqrt{6} \Phi_{test} \frac{\Delta\tilde r_c}{\tilde r_c} 
\end{equation}
\begin{equation}
\lesssim            \left \{\begin{array}{rl}
 5.97\times 10^{-11} & \textrm{(Solar system test)}, \\
 3.43\times 10^{-15} & \textrm{(Equivalence Principle test)}.
\end{array} \right .
\label{fifth-force}
\end{equation}
Therefore, in practice, for a particular model of $f(R)$ gravity, we just need to calculate $\psi_{out}$ to see whether it satisfies the fifth-force constraints \cite{Will:2014xja}, \cite{Capozziello:2007eu}. In the next section, we will discuss the curvature singularity problem for a particular form of $f(R)$ gravity model in combination with the fifth-force constraint.

\section{A General Model}\label{sec:genmodel}
In this section, we will consider a model proposed in \cite{Miranda:2009rs}:
\begin{equation}
f(R)= R + \alpha R_*\beta\left\{\left[1+\left(R/R_*\right)^n\right]^{-1/\beta}- 1\right\}~,
\label{fgenmodel}
\end{equation}
where $n$, $\beta$, $\alpha$ and $R_*$ are positive parameters of the model. $R_*$ is taken to be of the order of present day average curvature of the universe. The above function satisfies the condition $f(R)\rightarrow 0$ as $R\rightarrow 0$ and $f(R)\rightarrow R~ +$ constant, for large $R$. This is necessary for the correct GR limit in early cosmological epoch \cite{Starobinsky:2007hu}. Viable matter era demands $n > 0, \beta > 0$ or $n > 0, \beta < -n$ \cite{Amendola:2006we}. In this parametrisation, Starobinsky model corresponds to $n = 2$ \cite{Starobinsky:2007hu}, whereas Hu-Sawicki model can be obtained by plugging $\beta=1$ \cite{Hu:2007nk}. For the case of $n > 0$ and $\beta < 0$, the model becomes indistinguishable from the $\Lambda$CDM model \cite{Capozziello:2007eu}. In the original paper of \cite{Miranda:2009rs}, the model was analysed in the limit $\beta \rightarrow \infty$ and for $n = 1$, and it was shown that the model is free from curvature singularity issue. We will discuss about the model in this limit in the next section. In this section, we will analyse the model in its full generality given by Eq.~\eqref{fgenmodel}.

\begin{figure}
  \centering
   
   \includegraphics[width=0.45\textwidth]{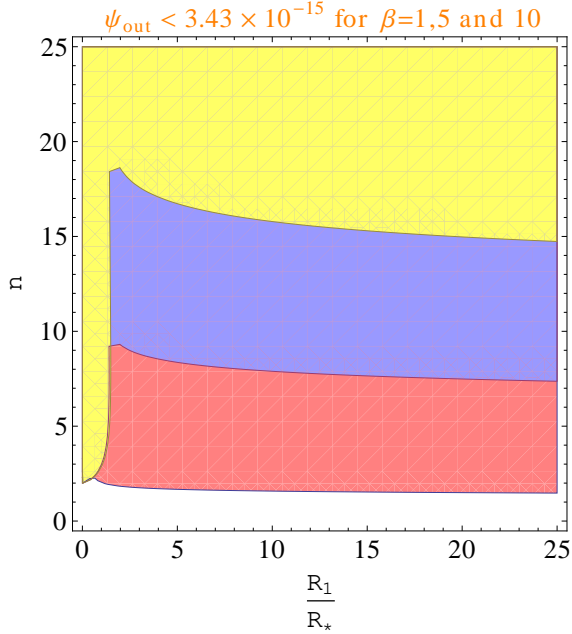} 
    
  \caption{parameter space of $n$ and $R_1/R_*$ for different values of $\beta$. Pink region shows allowed region for $\beta=1$, purple region for $\beta=5$ and yellow region for $\beta=10$.}
  \label{fig:fifthforce1}
\end{figure}

First, we briefly discuss the fifth-force constraints before moving into the curvature singularity issue which is the main focus of the article. In the limit of large curvature $R\gg R_*$, the form of $f(R)$ is reduced to
\begin{equation}
f(R)=R+\alpha R_*\beta\left\{\left(R/R_*\right)^{-n/\beta}-1\right\}~.
\label{fappxgenmoel}
\end{equation}
As discussed in the previous section, to evade the fifth-force, the canonical scalar field in the Einstein frame outside the test body must satisfy the constraint of Eq.~\eqref{fifth-force}. 
In the large curvature limit,
the scalar field at its minimum of the potential outside the test body is given by \cite{Thongkool:2009js}
\begin{equation}
\psi_{out} = \sqrt{\frac{3}{2\kappa^2}} \ln(1 + F_{,R}) \sim \frac{\sqrt6}{2\kappa}F_{,R} =-\frac{\sqrt 6}{2\kappa}n\alpha{\left(\frac{\kappa^2\rho_{out}}{R_*}\right)}^{-\frac{n}{\beta}-1}~,
\label{psioutcond}
\end{equation}
where we have used the minimisation condition of $V_{\text{eff}}$ in evaluating $R = \kappa^2\rho_{out}$ at the minimum of the effective potential. 
If $R_1$ is the curvature at the de Sitter minimum $R_1 = \kappa^2 \rho_{\text{crit}}$, we define a dimensionless variable $x_1=R_1/R_*$ to  obtain
\begin{equation}
\left|\psi_{out}\right| \sim \frac{\sqrt 6}{2 \kappa}n\alpha\left(\frac{ x_1\rho_{out}}{\rho_{\text{crit}}}\right)^{-\frac{n}{\beta}-1}~.
\label{psib}
\end{equation} 
Now  $\rho_{critical} \simeq 10^{-29} {\text gm/cm}^{3}$, and $\rho_{out} \simeq 10^{-24} gm/cm^{3}$ (typical baryonic/dark matter density). Because of $F(R) \ll f(R)$, $\alpha$ should be of order of unity. In this case, the fifth-force constraint of Eq.~\eqref{fifth-force} can be easily translated to the condition $n/\beta \gtrsim 2$ for $x_1 \sim \mathcal{O}(1)$. For general values of parameters the allowed regions are shown in Fig.~\ref{fig:fifthforce1}, where we have used the condition of de Sitter minimum in writing $\alpha$ in terms of other parameters. 
It is clear that the fifth force constraint is satisfied when $n/\beta\gtrsim2$ \cite{Thongkool:2009js}. It can be been seen that in the limit of  $\beta \rightarrow \infty$, the model immediately violates the fifth-force constraints \cite{Thongkool:2009js}, even though it can evade the curvature singularity problem \cite{Miranda:2009rs}. 

\subsection{Curvature Singularity: Static Analysis}
In this section, we will discuss the issue of curvature singularity in the Jordan frame for the model described by Eq.~\eqref{fgenmodel}. Looking at the form of the potential $V_J$, we argue for the presence of curvature singularity in this model. A preliminary analysis was done in \cite{Thongkool:2009js}, but in the Einstein frame. Moreover, the true nature of the singularity can be understood when the effects of the matter is taken care properly via $V_J^{\text{eff}}$ \cite{Frolov:2008uf}. Typically, the existence of matter pushes the minimum point close to the singular point. We will show it for `Log model' in the next section - See Fig.~\ref{fig:potentiallogmodel}. In the next subsection, we will analyse the dynamics of the field by solving the equation of motion to show that the field indeed reaches to the singular point in a time scale much smaller than the age of the Universe.  

\begin{figure}
  \centering
   $
   \begin{array}{c c}
   \includegraphics[width=0.5\textwidth]{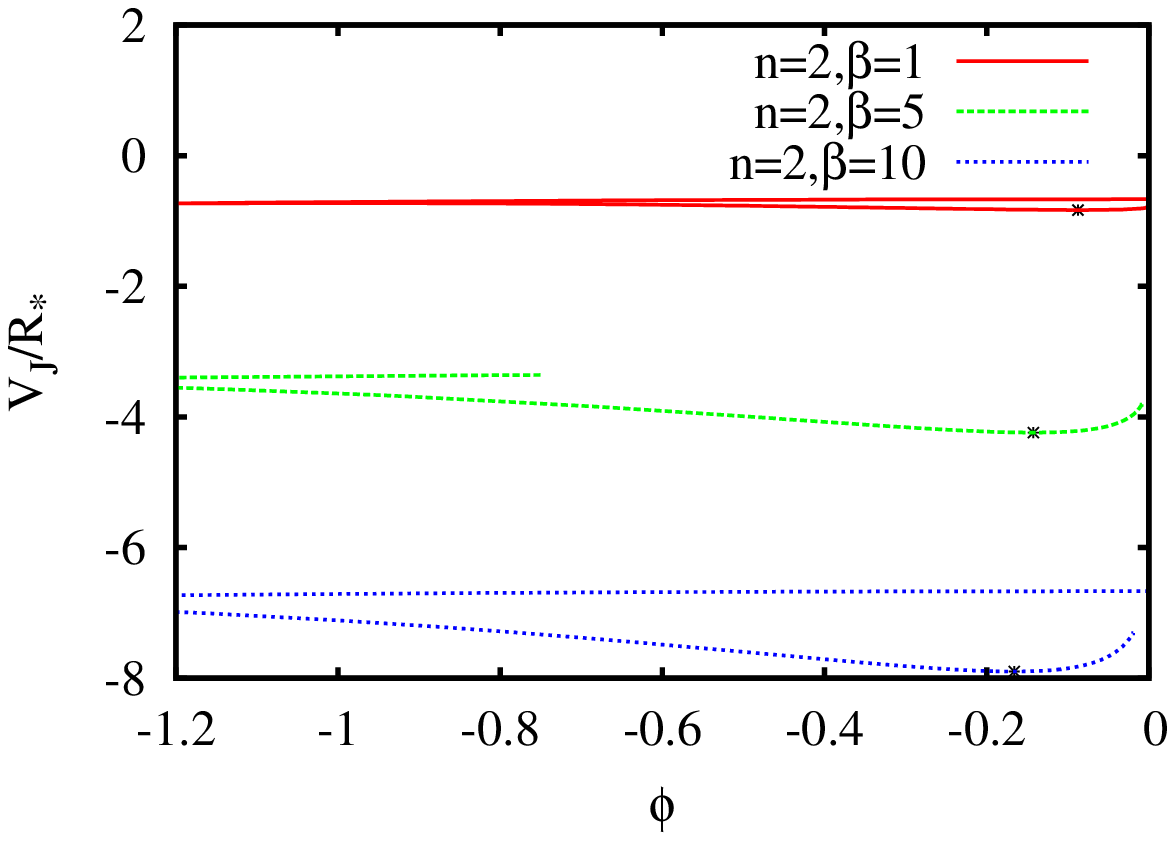} &
   \includegraphics[width=0.5\textwidth]{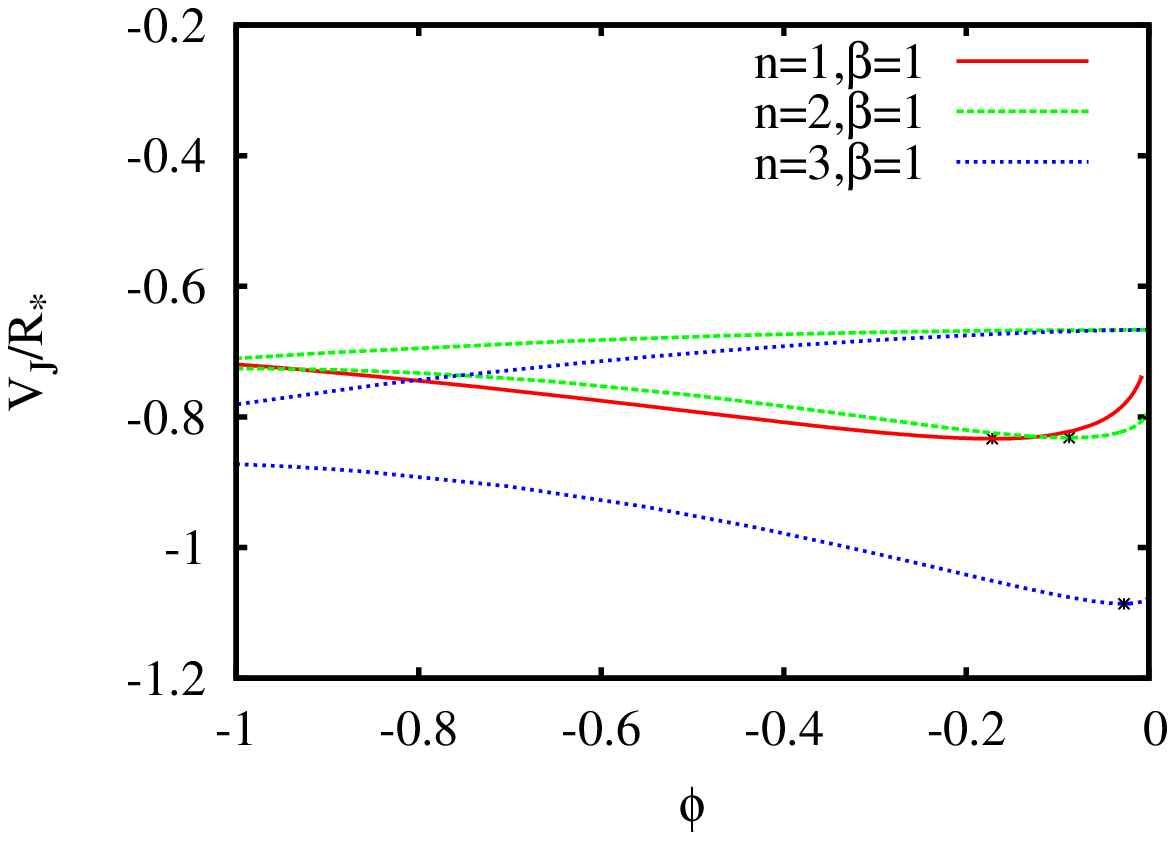}
   \end{array}
   $ 
  \caption{$V_J/R_*$ vs. $\phi$. In both figures $\alpha=2.0$. In the \emph{left} figure, $n=2$ and $\beta=1,5,10$ and in the \emph{right} figure $n=1,2,3$ and $\beta=1$. The minimum of a potential is marked with a dot sign.}
  \label{fig:potentialgenmodel}
\end{figure}

For our case, the scalar degree of freedom  $\phi$ in the Jordan frame can be identified as
\begin{equation}
\phi=-n\alpha\left(R/R_*\right)^{n-1}\left[1+\left(R/R_*\right)^n\right]^{-1/\beta-1}~. 
\label{phigenmodel}
\end{equation} 
Note that $R\rightarrow \infty$ when $\phi \rightarrow 0$, $i.e$ $\phi_{sing} = 0$ in this model. At the same time, we also note that positive values to $\phi$ correspond to negative curvature except for $n = 1$ case. 
In Fig.~\ref{fig:potentialgenmodel}, we plot the Jordon frame potential $V_J$ \footnote{The analytical expression is complicated, and is not very illuminating. Thus, we do not show it explicitly here.}. 	The potential is multivalued which is common to many $f(R)$ gravity models, and in this case the physically relevant potential is the lower one. 
 As described in the previous section, the de Sitter minimum of the potential, about which scalar field oscillates, can be obtained from the condition $dV_J/d\phi = 0$ which is equivalent to the condition $dV_E/d\psi = 0$.
We see from the plot that as we increase the value of $\beta$, the height of the potential barrier between the de Sitter point and the singularity point increases, and also the de Sitter minimum of the potential shifts away from the singularity in the field space. Thus, for a given $n$, larger values of $\beta$, the de Sitter minimum is at safer distance from the curvature singularity. This is in accordance with the findings of \cite{Thongkool:2009js} (in Einstein frame) where in the limit of $\beta \rightarrow \infty$, the potential barrier becomes infinitely large, thus inaccessible by the field and effectively hiding the singularity \cite{Thongkool:2009js, Miranda:2009rs}. On the other hand, looking at the right panel plot of Fig.~\ref{fig:potentialgenmodel} we see that making $n$ large pushes the minimum close to the singular point.  

\subsection{Curvature Singularity: Dynamical Analysis}

We consider a self gravitating system (e.g astronomical objects of dark matter cloud) whose mass density is changing with time. This is in contrast to the static configurations in $f(R)$ theories \cite{Kobayashi:2008tq, Babichev:2009fi, Upadhye:2009kt}. We solve Eq.~\eqref{diffaleq} for the contracting homogeneous and isotropic cloud of pressureless dust whose density $\rho_m$ is much greater than the critical density $\rho_{\text{crit}}$. For example, $\rho_{m}\sim 10^{-24}$ gm/cm$^3$ for a dust cloud in a galaxy. We parametrise the trace of energy-momentum tensor $T$ as
\begin{equation}
T (t) =-T_0\left(1+t/t_{ch}\right)~,
\label{Tgenmodel}
\end{equation}
where, $T_0=\rho_m$ is the energy-momentum density at $t=0$, and $t_{ch}$ is the characteristic time for density variations. 
It can be estimated by $t_{ch} \sim d/v$, where $d$ would be the typical dimension of the collapsing object and $v$ is its velocity during collapse. Here, we assume that the object collpases only under the effects of gravity and therefore velocity $v$ is nothing but the escape velocity on the surface of the object. Consequently, $t_{ch}$ can be calculated by $t_{ch}\sim \sqrt{3/(8\pi G\rho_m)}$ which comes out to be $\sim 1.34\times 10^{15}$ sec for $\rho_m\sim 10^{-24}gm/cm^3$. The time within which the system meets curvature singularity is denoted by $t_{sing}$. For a physical system, $t_{sing}$ must be smaller than the age of the Universe denoted by $t_U \sim 4\times 10^{17}$ sec. We have assumed that the gravitational field of the object is weak and hence the background metric is a flat Minkowski metric, and because of the homogeneity and isotropy D'Alembertian has been replaced with ordinary time derivative \cite{Reverberi:2012ew}. 
\begin{figure}
  \centering
   $
   \begin{array}{c c}
      \includegraphics[width=0.51\textwidth]{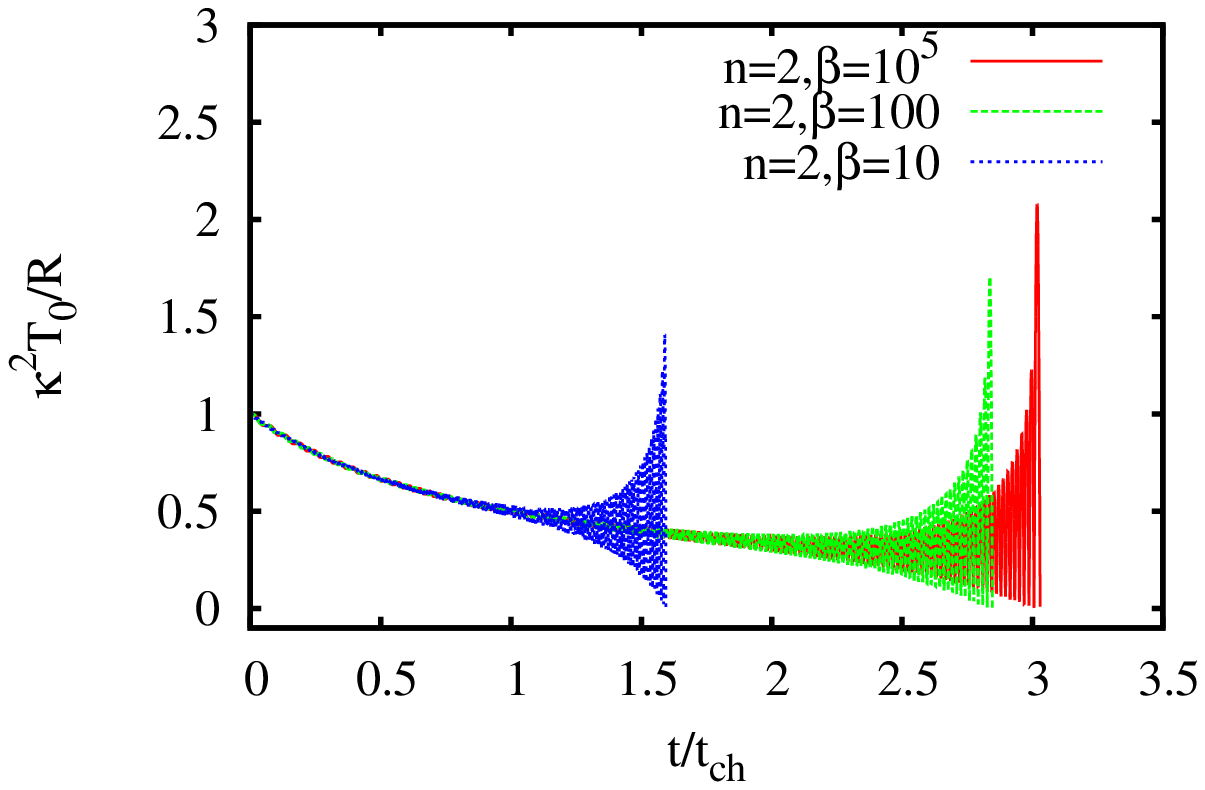}
   \includegraphics[width=0.51\textwidth]{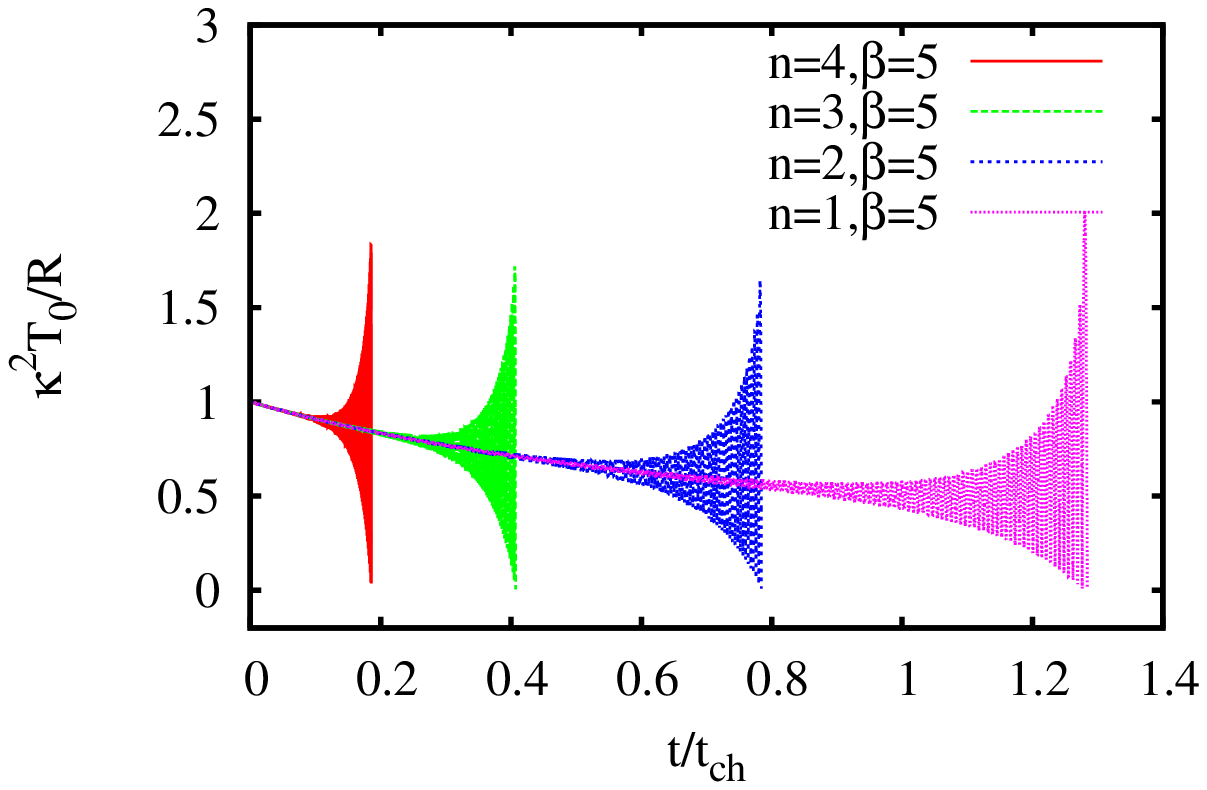} &
   \end{array}
   $ 
  \caption{The oscillation amplitudes are plotted against physical time for several parameter choices of the model. $\kappa^2T_0/R$ vs. $t$. In both the figures, $t_{ch}=1.34\times10^{15} sec$, $\eta=10^5$ and $\alpha=2$. 
  }
  \label{fig:soldyn1genmodel}
\end{figure}

With the above mentioned variation of matter density, the governing equation for the scalar field looks like
\begin{equation}
3\partial_t^2 F_{,R} +2F + R - R F_{,R} - \kappa^2 T_0\left(1+t/t_{ch}\right)=0~.
\label{eqmotion1}
\end{equation}  
Changing the variables from $R$ and $t$ to $y=\eta (R_*/R)$ and $\tau=t/t_{ch}$ we obtain
\begin{equation}
y''+\frac{n}{\beta}\frac{{y'}^2}{y}=\tau_{ch}^2\left[-y^{-n/\beta}\left(\left(1+\tau\right)-y^{-1}+\frac{2\alpha\beta}{\eta}\right)+n\alpha\eta^{-1-n/\beta}\left(1+\frac{2\beta}{n}\right)\right]~,
\label{dynofygenmodel}
\end{equation}
where, $\eta=\kappa^2T_0/R_*$, and $\tau_{ch}$ is given in terms of model parameters and characteristic time $t_{ch}$
\begin{equation}
\tau_{ch}=\left(\left(\kappa^2T_0/R_*\right)^{2+\frac{n}{\beta}}\frac{\beta}{n+\beta}\frac{R_*}{3n\alpha}\right)^{\frac{1}{2}}t_{ch}~.
\end{equation}
In deriving the above equation of motion, we have used the Eq.~\eqref{fappxgenmoel}. 
The singularity is reached when $y = \kappa^2 T_0/R$ becomes zero. The nonlinear behaviour of Eq.~\eqref{dynofygenmodel} can amplify the initial oscillation amplitude around the potential minimum and the field can reach to the singular point in a time scale that is smaller than the age of the Universe. The equation has been solved for $y'(0)=0$ and $y(0)=1$ initial values, and it is consistent with $R/R_* \gg1$ approximation. The solution of above equation for $t_{ch}=1.34\times 10^{15} Sec$ and different values of $n$ and $\beta$ is given in Fig.~\ref{fig:soldyn1genmodel}.

\begin{table}
\centering
\begin{tabular}{|l|l|l|l|}
  \hline
  $n$                          & $\beta$ & $n/\beta$ & $t_{sing}$(Sec)      \\
  \hline
    2                          & 1     & 2    & $8.6\times 10^{10}$ \\
      2                          & 2     & 1    & $6.9\times 10^{13}$     \\
        4                          & 5     & 0.8  & $2.5\times 10^{14}$      \\ 
  2                          & 5     & 0.4  & $1.0\times 10^{15}$      \\

  2                          & 100   & 0.02 & $3.9\times 10^{15}$      \\

  2                          & $10^5$  & $2\times 10^{-5}$ & $4.5\times 10^{15}$  \\
  \hline
\end{tabular}
\caption{$t_{sing}$ for different parameter values of the model defined by Eq.~\eqref{fgenmodel}. We have chosen $\alpha = 2$. For all these choices of parameters, the time scale of singularity is much smaller than the age of the Universe.}
\label{tabl:conditions}
\end{table}
It can be seen that the amplitude of the oscillations increase with time, and after a finite time the value of $y$ approaches zero which is a singular point. It is necessary that the time scale within which $y$ goes to the singular point must be shorter than $t_U$. As an example, taking $R_*\sim1/t_U^2$, $t_{sing}=2.5\times 10^{14}~Sec \ll t_{U}$  for $n=4$, $\beta=5$ and $\alpha=2$. 
It is clear from the plots that higher values of $\beta$ and lower $n$ values are needed for the considered model to remain free from curvature singularity. This is consistent with the primary idea that we had obtained by looking at the potential $V_J(\phi)$ for the scalar field. 
The time $t_{sing}$ has been listed in Table.~\ref{tabl:conditions} for different values of parameters $n$ and $\beta$. It is evident that the relevant parameter for the time scale of singularity $t_{sing}$ is the ratio between $n$ and $\beta$. The singularity time $t_{sing}$ increases with smaller values of $n/\beta$. 

Thus, it can be concluded that singularity can be avoided in principle by taking very large values of $\beta$ \cite{Miranda:2009rs} and/or small values of $n$. On the other hand, as discussed earlier, local gravity tests constrain the values of the parameters $n/\beta \gtrsim 2$. Therefore, it is nearly impossible practically to find a region in the parameter space where the model of Eq.~\eqref{fgenmodel} can be free from curvature singularity, as well as it can evade the local gravity tests.  

\section{`Log' Model}\label{sec:logmodel}
In the last section, we have seen that the curvature singularity can be avoided in principle by taking large $\beta$ and small $n$ limit of the general model given by \eqref{fgenmodel}. In \cite{Miranda:2009rs}, it has been claimed that if we take $n = 1$ and $\beta\rightarrow\infty$ limit of Eq.~\eqref{fgenmodel}, the generalised model is reduced to the following logarithmic function:
\begin{equation}
f(R)=R-\alpha R_*~ln\left(1+R/R_*\right)~.
\label{flogmodel}
\end{equation}
Subsequently, this particular form of the model has been analysed and it is shown that the curvature singularity is absent \cite{Miranda:2009rs}, even though it violates the fifth force constraint of $n/\beta \gtrsim 2$. First of all, this reduction is only possible under the assumption $R \ll R_*$ and it breaks down immediately for the analysis of curvature singularity. Therefore, we consider Eq.~\eqref{flogmodel} as an independent model and revisit the issue of curvature singularity in the presence of matter.

\begin{figure}
  \centering
   $
   \begin{array}{c c}
   \includegraphics[width=0.5\textwidth]{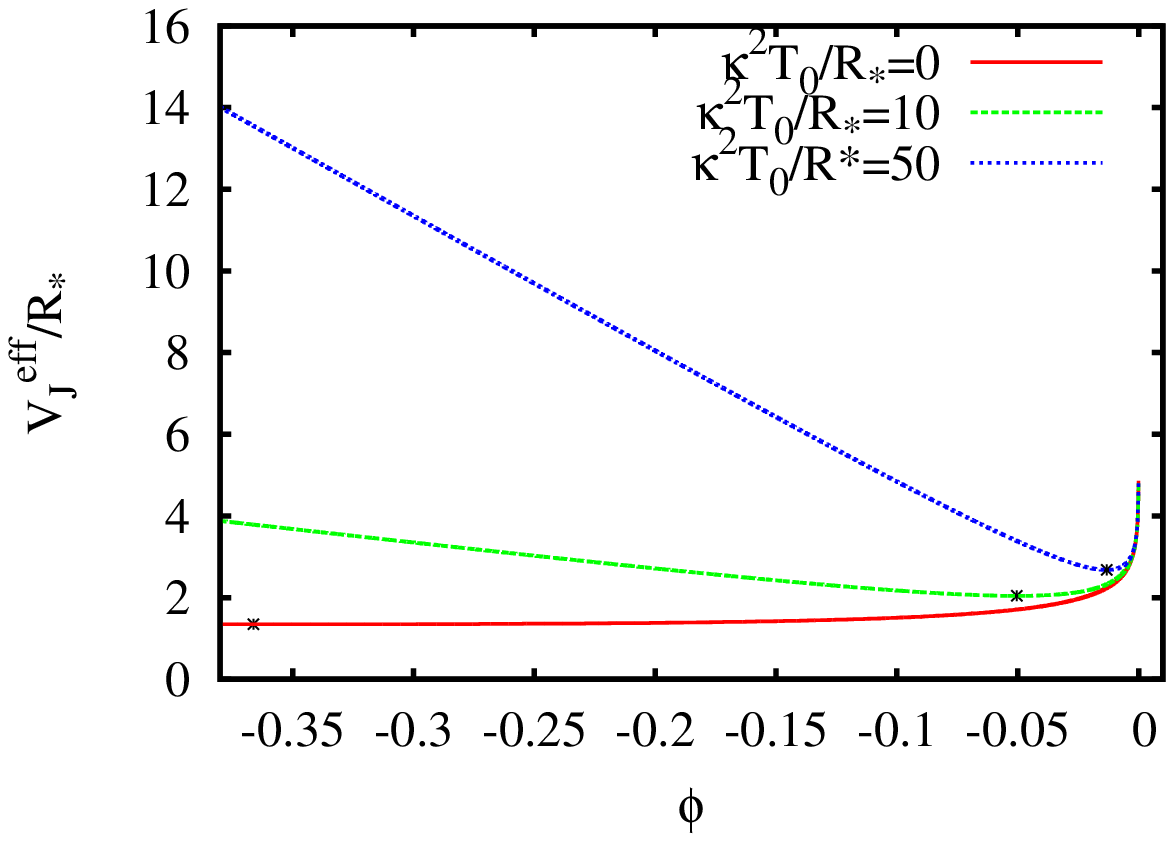} & 
   \includegraphics[width=0.51\textwidth]{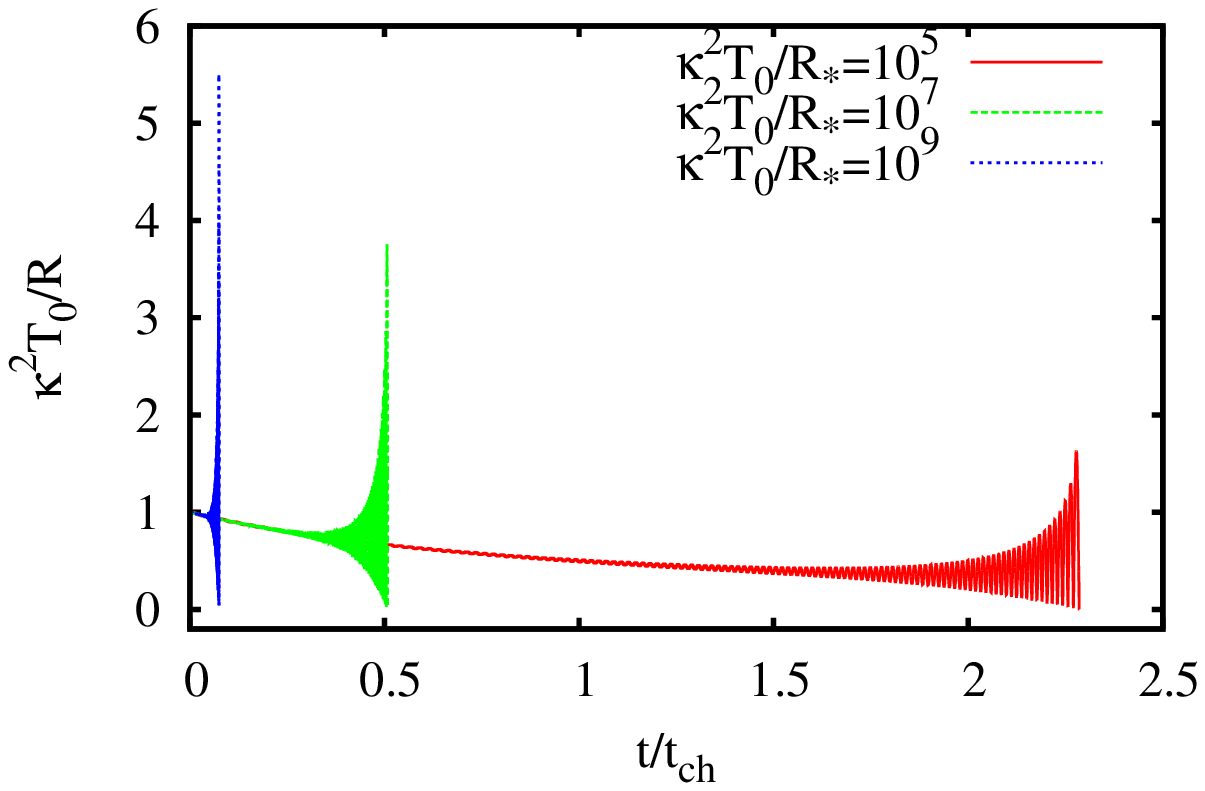}
   \end{array}
   $ 
  \caption{Left: $V_J^{\text{eff}}$ vs. $\phi$ where dots represent the potential minimum. In the figure, $\kappa^2T_0/R_*=0,10,50$ and $\alpha=1.5$. Right: $\kappa^2T_0/R$ vs. $t/t_{ch}$ for $\eta=10^5,10^7,10^9$ with $\alpha=1.5$.  
  }
  \label{fig:potentiallogmodel}
\end{figure}

The scalar field $\phi$ in the Jordan frame for the model can be expressed as
\begin{equation}
\phi=-\frac{\alpha}{1+R/R_*}~,
\end{equation}
and we see that $\phi\rightarrow 0$ corresponds to infinite $R$ corresponding to the  curvature singularity. To investigate the issue, we first find out the potential
\begin{equation}
V_J(\phi)/R_*=-\frac{1}{3}\phi-\frac{1}{3}\alpha\phi+\frac{1}{6}\phi^2-\frac{2}{3}\alpha\phi \ln{\left(-\alpha/\phi\right)}+\frac{1}{3}\alpha\ln{\left(-1/\phi\right)}~,
\label{potential1logmodel}
\end{equation}
and note that $V_J(\phi)\rightarrow\infty$ as $\phi$ reaches the singularity point. Naively, the presence of the infinite potential barrier was the argument in claiming the absence of curvature singularity \cite{Thongkool:2009js}, \cite{Miranda:2009rs}. But including the matter source in term of a collapsing cloud of Eq.~\eqref{Tgenmodel}, the effective potential can be written as 
\begin{equation}
\frac{V_{J}^{\text{eff}}(\phi,t)}{R_*}=-\frac{1}{3}\phi-\frac{1}{3}\alpha\phi+\frac{1}{6}\phi^2-\frac{2}{3}\alpha\phi \ln{\left(-\frac{\alpha}{\phi}\right)}+\frac{1}{3}\alpha\ln{\left(-\frac{1}{\phi}\right)}-\frac{1}{3}\frac{\kappa^2T_0}{R_*}\phi\left(1+\frac{t}{t_{ch}}\right)~.
\label{potential2logmodel}
\end{equation}
For different values of $\kappa^2T_0/R_*$ at $t=t_{ch}$, the plot of $V_{J}^{\text{eff}}$ vs. $\phi$ has been shown in Fig.~[\ref{fig:potentiallogmodel}]. The de Sitter points are marked by stars in the plot. From the figure, it can be seen that the minimum of the potential shifts towards the singularity ($\phi=0$) as the density of matter increases \cite{Frolov:2008uf}. One can conclude that the possibility that the scalar field $\phi$ may hit the curvature singularity is more likely in the presence of matter than in the vacuum. But, it requires the understanding of dynamics of the field which we will analyse now.

We investigate the presence of curvature singularity by solving Eq.~\eqref{diffaleq} numerically in contracting astrophysical objects as we have done in the previous section. The equation of motion in this case reads 
\begin{equation}
{y}''-\tau_{ch}^2\left[2 ln(y)- 2ln(\eta) + \frac{\eta}{\alpha}\frac{1}{y}+1-\frac{\eta}{\alpha}\left(1+\tau\right)\right]=0~,
\label{fulldiffeqlogmodel}
\end{equation}
where the approximations at large curvature limit 
$F(R)\simeq-\alpha R_*ln\left(R/R_*\right)$ and $F_{,R}\simeq- (\alpha R_*)/R$ have been used.
Here, $\tau_{ch}$ is given by
\begin{equation}
\tau_{ch}=\sqrt{\frac{\eta R_*}{3}}t_{ch}.
\end{equation} 
The solution of the above equation $y(\tau)$ is plotted in Fig.~\ref{fig:potentiallogmodel} for $\rho_m=10^{-24},10^{-22}$ and $10^{-20} gm/cm^3$ i.e. for $\eta=10^5, 10^7$ and $10^9$. The characteristic time $t_{ch}$ for these densities are $1.34\times 10^{15}, 1.34\times 10^{14}$ and $1.34\times 10^{13}$ sec correspondingly. The initial conditions have been chosen as $y(0)=1$ and $y'(0)=0$. It is seen that the oscillations of $y$ grows with time and eventually meets the singularity. It is evident from the figure that the singularity is reached earlier for larger $\eta$. It was noted earlier that the model is singularity free in the absence of matter because of its large potential barrier at the singular point. But we show that in the presence of astrophysical densities in a collapsing object the system indeed can reach to the singular point.  

\section{Discussion and Conclusion}\label{sec:conclusion}
In this work, we have investigated the existence of curvature singularity problem in $f(R)$ model given by Eq.~\eqref{fgenmodel}  and  by Eq.~\eqref{flogmodel} \cite{Miranda:2009rs}. For our analysis we have chosen this particular model as this form of $f(R)$ incorporates several $f(R)$ models well known in the literature. We have made the dynamical analysis of the fluctuations of the associated scalar field around the minimum of the potential to show that the field indeed reaches to the singular point when the effects of the matter are considered. 

We have carried out our analysis of curvature singularity in the Jordan frame where the theory has been rewritten in terms of a scalar field $\phi$. The singularity is reached when $\phi=\phi_{sing}$, and in our case, $\phi_{sing}=0$. The scalar field oscillates around its de Sitter minimum of the potential $V_J$. The potential barrier at the singularity point, and the distance between the de Sitter point and the singular point are finite. Therefore the scalar field can reach to $\phi_{sing}$ during its cosmological evolution or during the collapse of dense astrophysical objects. In this work, we have shown this effect by solving the nonlinear evolution equation of the field. With increasing value of $n/\beta $ in the considered model, the system reaches to the singularity earlier. In particular, we have solved the stress equation for collapsing astrophysical object and shown that the Ricci scalar $R$ oscillates and after a time $ t \sim t_{sing}$, it reaches to a divergent value. The time within which singularity is reached, $i.e.$ $t_{sing}$, is less than the age of the Universe for generic values of $n$ and $\beta$. On the other hand, local gravity tests put the constraint on the model parameters such that $n/\beta \gtrsim 2$. Taking large enough values of $\beta$ can in principle help us avoiding the singularity in cosmological time scale, but it immediately violates the fifth force constraints. Therefore, we conclude that the model can not satisfy constraints coming from the curvature singularity and fifth force simultaneously.\\
On the other hand, the `Log Model' as discussed in Sec.~\ref{sec:logmodel} seems to be free from curvature singularity \cite{Thongkool:2009js},\cite{Miranda:2009rs}. But, this conclusion was drawn in the absence of matter. When we incorporate the effects of matter in the calculations, we see that the minimum of the effective potential moves closer to the singular point and the height of the barrier at $\phi_{sing}$ becomes smaller. In fact, the solutions of the scalar field dynamics show that the singularity is reached in a time scale smaller than the age of the Universe.  

Looking at the present analysis in conjunction with previous understanding, it is tempting to think that it is very difficult, if not impossible, to construct any function $f(R)$ that simultaneously satisfies both fifth force constraint  and  resolves the singularity problem. As we have seen, and also has been noted earlier \cite{Thongkool:2009js}, the height of the potential barrier goes opposite to satisfying the fifth force constraint. For example, in the considered model, the height of the barrier (thus avoidance of singularity) is proportional to $\beta$, but having large $\beta$ immediately violates the fifth force constraints. Therefore, it would be very interesting in finding generic structure of the function $f(R)$ that can simultaneously satisfy both the above mentioned constraints. We hope to look into this issue in future. 

\section*{Acknowledgements}
KD is partially supported by a Ramanujan Fellowship and Max Planck Society-DST Visiting Fellowship. SP and AP would like to thank Theory Division of SINP where part of the work has been carried out as visitors. 
\appendix

\end{document}